\documentstyle[aps,prb,epsf,multicol]{revtex} 
\input epsf
\newcommand{\dir}{Figs}

\newcommand{\fig}[4]
{
     \noindent
     \unitlength=1mm
     \begin{picture}(#2,#3)
     \put(0,0){\leavevmode \epsfxsize=#2mm \epsffile{\dir/#1}}
     \end{picture}
   \noindent
#4
}
\newcommand{\rr}{ {\bf r} }
\newcommand{\ru}{ {\bf \hat{r}} }

\newcommand{\nn}{ {\bf n} }
\newcommand{\uu}{ {\bf u} }
\newcommand{\QQ}{ {\bf Q} }
\newcommand{\II}{ {\bf I} }

\begin{document} 

\newcommand{\CCdtotal}
{
\caption{
Total density as a function of the distance $z$ from the wall for 
different grafting densities $\Sigma$. The curves are shifted by 
$0.2/\sigma_0^3$ for better visibility. Grafting densities are 
(from top to bottom) $\Sigma \:\sigma_0^2 =$ 0.563, 0.444, 
0.25, 0.174, 0.111, 0.0625, 0.0278, 0.
}
\label{fig:dtotal}
\bigskip
}

\newcommand{\CCdexc}
{
\caption{
Excess density vs. grafting density $\Sigma$.
}
\label{fig:dexc}
\bigskip
}

\newcommand{\CCdsolv}
{
\caption{
Density of solvent particles as a function of the distance $z$ from the 
wall for different grafting densities: (from bottom to top) 
$\Sigma \: \sigma_0^2 =$ 0.563, 0.444, 0.34,
0.25, 0.174, 0.111, 0.0625, 0.0278, 0.00694, 0. Inset shows 
a blow-up of the same data.
}
\label{fig:dsolv}
\bigskip
}

\newcommand{\CCdchain}
{
\caption{
Density of chain monomers as a function of the distance $z$ from the 
wall for different grafting densities: (from top to bottom) 
$\Sigma \: \sigma_0^2 =$  0.563, 0.444, 0.34,
0.25,  0.174, 0.111, 0.0625, 0.0278, 0.00694. 
Inset shows the chain density distributions, i. e., the 
normalized density profiles $\rho(z)/\int dz' \: \rho(z') $.
}
\label{fig:dchain}
\bigskip
}

\newcommand{\CCdends}
{
\caption{
Distribution of chain end monomers vs. distance $z$ from the surface
for different grafting densities: (following the arrows) 
$\Sigma \: \sigma_0^2$ = 0.563, 0.444, 0.34,
0.25,  0.174, 0.111, 0.0625, 0.0278, 0.00694. 
Inset shows the same curves on a semi-logarithmic scale.
}
\label{fig:dends}
\bigskip
}

\newcommand{\CCtchains}
{
\caption{
Average tilt angle~$\langle \theta \rangle$ of the head-to-end 
vector of the chains as a function of the grafting density $\Sigma$
without solvent (open symbols) and with solvent (filled symbols).
}
\label{fig:tchains}
\bigskip
}

\newcommand{\CCtparticles}
{
\caption{
Average tilt angle~$\langle \theta \rangle$ of particles
as a function of the grafting density $\Sigma$.
Open symbols: Chain monomers, Filled symbols: solvent particles.
}
\label{fig:tparticles}
\bigskip
}

\newcommand{\CCthist}
{
\caption{
Distribution $P(\cos\theta)$ of tilt angles for solvent particles
(left) and chain monomers (right) and different grafting
densities (following the arrows): $\Sigma \: \sigma_0^2$ =  
0.694, 0.563, 0.444, 0.34, 0.25,  0.174, 0.128, 0.111, 
0.0625, 0.0278, 0.00694, and 0 (only left figure).  
The distribution for $\Sigma \: \sigma_0^2 = 0.128$ is
shown as dashed line.
}
\label{fig:thist}
\bigskip
}

\newcommand{\CCthistmax}
{
\caption{
Position of the maximum angle $\theta_{max}$ vs. grafting 
density $\Sigma$ for chain particles (open squares) and
solvent particles (filled circles). 
}
\label{fig:thistmax}
\bigskip
}

\newcommand{\CCprofop}
{
\caption{
Profiles of the order parameter $\langle S \rangle$ (thick lines) and the 
components of the director $\nn$ (thin lines) vs. $z$ in the 
film for different grafting densities $\Sigma$ as indicated.
thin solid line: $\langle | n_z | \rangle$; 
thin short dashed line: $\langle |n_x| \rangle $;
thin long dashed line: $\langle | n_y | \rangle$.
}
\label{fig:profop}
\bigskip
}

\newcommand{\CCnzfilm}
{
\caption{
$z$ component of the director in the middle of the film
as a function of the grafting density.
}
\label{fig:nzfilm}
\bigskip
}

%

\title{Surface anchoring on layers of grafted liquid-crystalline chain
  molecules: A computer simulation}

\author{Harald Lange$^{\dag, \ddag}$ and Friederike Schmid$^{\dag}$}

\address{
$\dag$ Fakult\"at f\"ur Physik, Universit\"at Bielefeld, 
         33615 Bielefeld, Germany \\
$\ddag$ Institut f\"ur Physik, Universit\"at Mainz, 
55099 Mainz, Germany
}

\setcounter{page}{1}
\maketitle 
\tighten

\begin{abstract}

By Monte Carlo simulations of a soft ellipsoid model for liquid crystals, 
we study whether a layer of grafted liquid-crystalline chain molecules
can induce tilt in a nematic fluid. The chains are fairly short (four
monomers) and made of the same particles as the solvent. They are 
attached to a substrate which favors parallel (planar) alignment. 
At low grafting densities, the substrate dominates and we observe
planar alignment. On increasing the grafting density, we find a first
order transition from planar to tilted alignment. Beyond the transition,
the tilt angle with respect to the surface normal decreases continuously.
The range of accessible anchoring angles is quite large. 

\end{abstract}

\begin{center}
\end{center}

%
%

\section{Introduction} 
\label{sec:intro}

\begin{multicols}{2}

Surfaces orient nematic liquid crystals, because they align nearby 
particles~\cite{degennes,chandrasekhar,jerome}. This phenomenon, called 
surface anchoring, is not only interesting from a fundamental point 
of view~\cite{cognard,abbott,mike}; it also plays a key role in liquid crystal 
display devices~\cite{bahadur,schadt}. Surface modifications which allow 
to tune the anchoring properties of a surface -- in particular, the anchoring 
angle -- have great practical importance~\cite{scheuble,patel}.

To be useful in technological applications, such modified surfaces must 
be physically and chemically stable. Therefore, preparation techniques where 
molecules are attached chemically to the surface are desirable. The coupling 
between the surface and the nematic fluid must be well-defined, i.~e., strong. 
This calls for the use of large molecules (polymers), which have themselves 
liquid crystalline properties. 

Based on such considerations, Peng, Johannsmann and R\"uhe have recently 
investigated the anchoring behavior of nematic liquid crystals on layers 
of grafted side-chain liquid crystalline polymers~\cite{peng1,peng2,peng3,peng4}. 
A ''grafting-from'' technique developed by R\"uhe et al~\cite{ruehe,prucker} 
allowed to synthesize brushes of up to 230 nm dry thickness. They were
forced to grow into a homogeneously aligned structure by an appropriate 
treatment of the substrate prior to the grafting step. When brought in 
contact with a nematic fluid, they transfered their orientation into the 
fluid~\cite{peng3}. Thus Peng et al have demonstrated experimentally that 
surfaces with grafted liquid crystalline chain molecules can align
nematic liquid crystals.

The idea of using liquid crystalline polymer brushes as alignment layers 
goes back to Halperin and Williams~\cite{avi1,avi2,avi3}. They suggested to 
create a competition between the orienting effect of the bare substrate, 
and that of stretched chains. The anchoring angle should then be determined by 
the properties of the substrate at low grafting densities, and by the chain 
stretching at high grafting densities. Simple mean field considerations lead 
Halperin and Williams to the conclusion that the two regimes should be 
separated by a second order transition. This implies that arbitrary anchoring 
angles can be realized at intermediate grafting densities.

The actual calculation of Halperin and Williams relies on two assumptions: 
The chains are very long and can be treated as entropic springs, and they 
are swollen by a good nematic solvent of low molecular weight. At least the 
second condition is not met in real systems: The brushes of Peng et al 
were only swollen to a very limited extent, even in a chemically similar 
solvent.  Nevertheless, the general idea still seems worth pursuing:
Consider a substrate which is decorated with liquid-crystalline chain 
molecules (main-chain liquid-crystals for simplicity), and which favors 
parallel alignment: Without grafted chains, the substrate aligns the fluid 
in a planar way. At high grafting densities, the chain molecules are forced 
to stand up. The question is: What happens at intermediate densities?

The high grafting density limit has been discussed in an earlier 
paper~\cite{harald1} for the case of short chains in a moderate solvent. 
Based on a simple theoretical model, we have conjectured that there exists
a continuous anchoring transition from tilted to perpendicular anchoring, 
which is driven by the grafting density. The theory predicts that the 
anchoring angle (with respect to the surface normal) is always smaller than 
the tilt angle inside of the chain layer. Thus the anchoring transition in 
the bulk preempts the corresponding tilting transition in the chain layer, 
and may even exist if there is no such tilting transition. Computer simulations 
were consistent with that picture, though not conclusive. This anchoring 
transition has nothing to do with the transition predicted by Halperin
and Williams, which separates a phase with planar alignment from a 
phase with tilted alignment. 

From a practical point of view, the low grafting density regime is more 
important, since low grafting densities can be realized much more easily 
in experiments. In the present paper, we study the anchoring behavior at
low and intermediate grafting densities by Monte Carlo simulations of an 
idealized, coarse-grained model. The general setup of the model is such that, 
strictly speaking, the Halperin/Williams theory does not apply: The chains 
are much too short. Nevertheless, we will see that the anchoring transition 
persists -- now as a weakly first oder transition. Our results demonstrate 
that the scenario predicted by Halperin and Williams is valid in a much wider 
range of parameters than that for which it was derived originally. 

In the next section, we introduce the model and describe the simulation 
method.  The results are presented in section \ref{sec:results}. 
We summarize and discuss our findings in section \ref{sec:summary}.

\section{Model and Method} 
\label{sec:simulation}

In our model, both the chain monomers and the solvent particles
are represented by soft ellipsoidal particles with elongation 
$\kappa = \sigma_{\mbox{\tiny end-end}}/\sigma_{\mbox{\tiny side-side}}=3$.
Two particles $i$ and $j$ with orientations $\uu_i$ and $\uu_j$ 
($|\uu| = 1$) at the positions $\rr_i$ and $\rr_j$ interact with 
the pair potential 
\begin{equation}
\label{eq:vij}
V_{ij}
= \left\{ \begin{array}{lcr}
4 \epsilon_0 \: (X_{ij}^{12} - X_{ij}^{6}) + \epsilon_0 & : & X_{ij}^6 > 
1/2 \\
0 & : & \mbox{otherwise}
\end{array} \right. ,
\end{equation}
where $X_{ij} = \sigma_0/(r_{ij}-\sigma_{ij}+\sigma_0)$ 
depends on the center-center distance $r_{ij} = |\rr_i - \rr_j|$
and the contact distance $\sigma_{ij}$ of two ellipsoids in 
the direction $\ru_{ij} = (\rr_i - \rr_j)/r_{ij}$. 
We approximate the latter by the gaussian overlap function 
suggested by Berne and Pechukas~\cite{berne}
\begin{eqnarray}
\lefteqn{\sigma_{ij}(\uu_i,\uu_j,\ru_{ij})
= \sigma_0 \:
\Big\{ 1 \: - \frac{\chi}{2}}  
\nonumber \\&& \quad
 \Big[
\frac{(\uu_i\cdot\ru_{ij}+\uu_j\cdot\ru_{ij})^2}
     {1+\chi \uu_i\cdot\uu_j}
+ 
\frac{(\uu_i\cdot\ru_{ij}-\uu_j\cdot\ru_{ij})^2}
     {1-\chi \uu_i\cdot \uu_j}
\Big] \Big\}^{-1/2},
\end{eqnarray}
with $\chi=(\kappa^2-1)/(\kappa^2+1)$. The interactions 
(\ref{eq:vij}) are identical for solvent particles and chain monomers. 
In addition, chain monomers are connected by bonds of length $b$, which 
are subject to a spring potential with an equilibrium length $b=b_0$ 
and a logarithmic cutoff at $|b-b_0|=b_s$.
\begin{equation}
\label{eq:fene}
V_{S}(b) = \left\{ \begin{array}{l c r}
\displaystyle - \frac{k_s}{2} b_s^2 \ln\Big( 1 - \frac{(b-b_0)^2}{b_s{}^2} \Big)
&: & |b-b_0| < b_s \\
\infty &: & |b-b_0| > b_s
\end{array} \right. .
\end{equation}
The chains are stiff by virtue of a stiffness potential
\begin{equation}
\label{eq:ba}
V_{A} (\theta_1,\theta_2,\theta_{12}) = 
- k_a \; [ \: \cos(\theta_1) + \cos(\theta_2) + 2 \cos(\theta_{12}) \: ],
\end{equation}
which depends on the angles $\theta_1$ and $\theta_2$ between 
the orientation of a monomer and the adjacent bonds, and the
angle $\theta_{12}$ between the two bonds.

The system is confined between two hard walls at $z=0$ and $z=L_z$.
A hard core potential prevents the particles from penetrating the 
walls,
\begin{equation}
\label{eq:wall}
V_{W}(z) = \left\{ \begin{array}{l cc }
0 & :& \quad d_z(\theta) < z < L_z - d_z(\theta) \\
\infty & :& \quad \mbox{otherwise} 
\end{array} \right. ,
\end{equation}
\begin{equation}
d_z(\theta) = \sigma_0/2 \: \sqrt{1+\cos^2 (\theta) \: (\kappa^2-1)}.
\end{equation}
The function $d_z(\theta)$ 
is the contact distance between a surface and an ellipsoid of 
diameter $\sigma_0$ and elongation $\kappa$, with the long axis 
oriented at the angle $\theta$ with respect to the surface normal.
The first bond of each chain is attached to one of the surfaces.
The grafting points are on a regular square lattice. The grafting
is fully flexible, i. e., no energy contribution couples to the
angle between the first bond and the surface normal.

We chose the model parameters  $k_s=10 \epsilon_0/\sigma_0^2$,
$k_a=10 \epsilon_0$, $b_0=4\sigma_0$ and $b_s=0.8\sigma_0$. 
The simulations were performed at the temperature $T=0.5 \epsilon_0/k_B$ 
and pressure $P=3 \epsilon_0/\sigma_0^3$. This corresponds to a 
state well in the nematic phase: The transition to the isotropic phase 
occurs at the pressure~\cite{harald2,harald3} $P=2.3 \epsilon_0/\sigma_0^3$. 
The bulk number density was $\langle \rho \rangle = 0.313/\sigma_0^3$.
The length of the chains was four bonds, (i. e., four monomers plus 
the grafting point). They are very stiff: The persistence length
is roughly 210 $\sigma_0$, and the end-to-end distance of free
chains is 14.8 $\sigma_0$. Compared to experimental systems, 
our chains correspond to main chain oligomers with very few 
(less than ten) mesogenic units. We have not attempted any 
quantitative mapping on a specific system.

The simulations were conducted in the $NPT$ ensemble, at constant 
temperature and pressure, and a fixed number of particles. 
The systems contained roughly 2000 solvent particles (the numbers
varied slightly in the different runs) and a varying number of chains 
with four monomers. The simulation boxes were rectangular with fixed 
lateral size $L_{\parallel}$ and fluctuating length $L_z$. The lateral
size was kept fixed in order to maintain a constant grafting density
$\Sigma$. It was chosen $L_{\parallel} = 12 \sigma_0$ in most runs,
except for four runs which sampled grafting densities between 
$\Sigma=0.111/\sigma_0^2$ ($4\times4$ chains on an area $12\cdot 12
\sigma_0^2$) and $\Sigma = 0.174/\sigma_0^2$ ($5 \times 5$ chains). 
These were conducted in systems with $5 \times 5$ chains and slightly 
larger surface areas. The length $L_z$ of the boxes was roughly
$L_z \approx 45-64 \sigma_0 (\pm 10 \sigma_0)$ (higher for higher
grafting density. Monte Carlo moves were carried out in random order 
and included particle displacements, particle rotations, and rescaling 
of the simulation box in the $z$ direction. They were accepted or rejected 
according to a Metropolis prescription with the effective Hamiltonian 
\begin{equation}
H = E + P L_z - N T \log(L_z).
\end{equation}
\end{multicols}\twocolumn

\begin{figure}[t]
\noindent
\fig{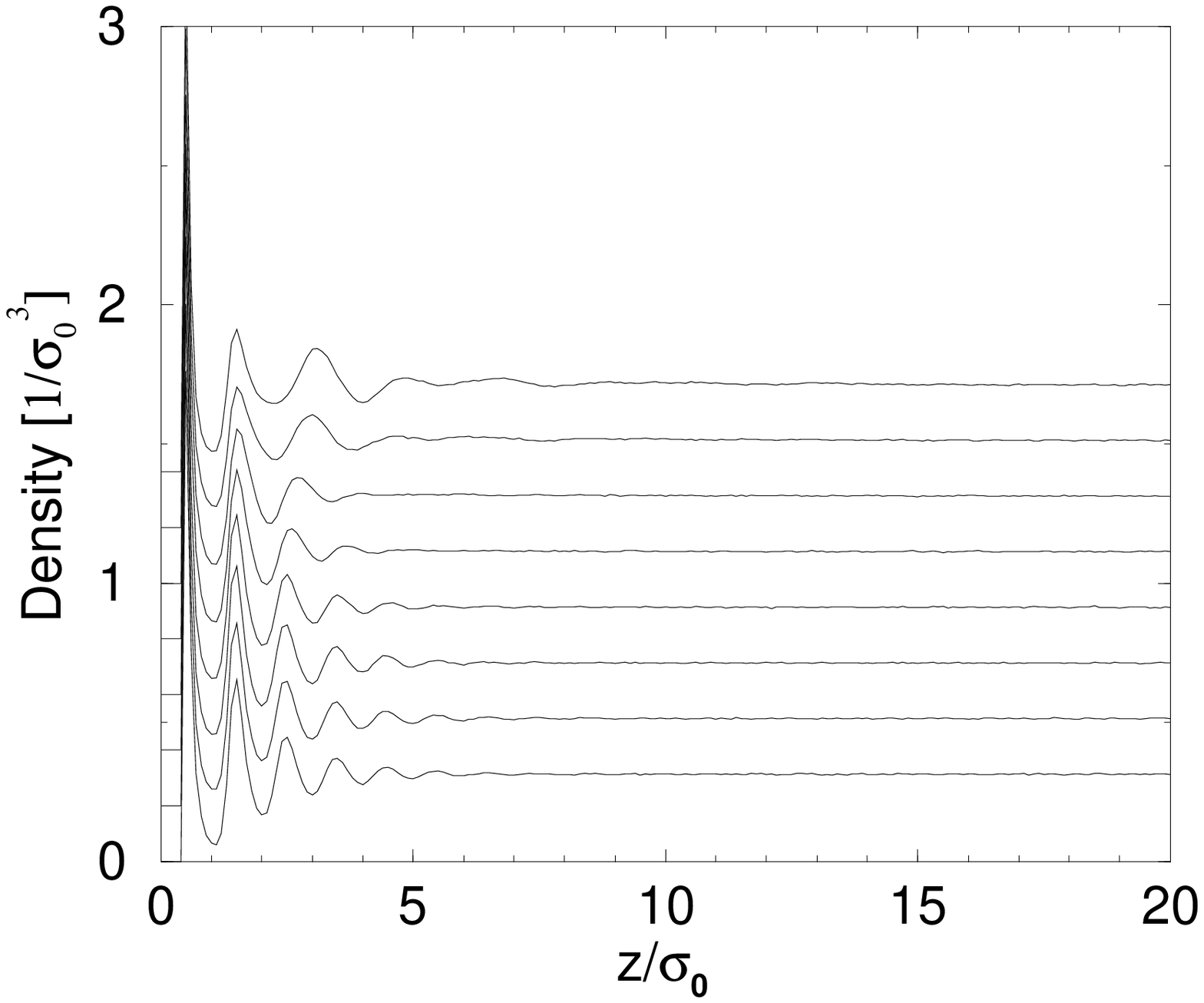}{80}{70}{}
\CCdtotal
\end{figure}
\noindent

Here $E$ is the internal energy and $N$ the total number of ellipsoids 
in the system (solvent particles and chain monomers). In addition, we 
have also implemented special Configurational Bias Monte Carlo 
moves~\cite{rosenbluth,frenkel}, which are specially adapted for our 
problem: The bonds of a randomly chosen chain were removed and 
redistributed such that a new chain was grown from the solvent. 
We calculated the probability $P_{new}$ of constructing this
particular new chain, and the probability $P_{old}$ of just 
reconstructing the same old chain. The new chain was then accepted 
with the probability
\begin{equation}
W_{\mbox{\tiny old} \to \mbox{\tiny new}} =
\min \big( 1, \frac{P_{\mbox{\tiny old}}}{P_{\mbox{\tiny new}}}
\exp (- \Delta E/T) \big),
\end{equation}
where $\Delta E$ is the energy difference between the old and the new chain. 
Details of this algorithm have been published elsewhere~\cite{harald3}. 
One ``Monte Carlo step'' consists on average of $N$ attempted translations, 
2 $N$ attempted rotations, one attempt of rescaling the box, and one 
configurational bias move.

Preliminary runs which started from initial configurations with particles 
oriented in the $z$ direction showed clearly that the walls align
the solvent particles in a parallel way. The particles close to the walls 
turned around first, and the parallel orientation then propagated 
into the middle of the film. Unfortunately, this usually lead to a state 
where the director at the two walls pointed into two different directions, 
and was forced to twist around in the middle of the cell. In the later
runs, we therefore prepared the configurations such that the particles 
all pointed into the $x$-direction initially. To this end, we applied a 
strong orienting field during roughly 50.000 Monte Carlo steps. 
After turning this field off, the system was equilibrated over at least 
1 million Monte Carlo steps. Data were then collected over 5 million or 
more Monte Carlo steps. 

Due to the complexity of the system, the
calculations were very time consuming: Every run required several
months on a Pentium III 800 processor. Therefore, we have not yet
attempted to study larger systems. Unfortunately, this limits
among other the maximum chain length -- one has to ensure that
a chain free ``bulk region`` remains in the middle of the cell.
Simulations of systems with longer grafted chains will clearly
be interesting in the future.

\section{Results} 
\label{sec:results}

Figure \ref{fig:dtotal} shows profiles of the total density
(chain monomers and solvent particles) near the wall.
They exhibit strong layering close to the wall, with decaying
amplitude further away from the wall. The position of the first
peak is independent of the grafting density. The distances between
the other peaks increase with increasing grafting density.
This is a first indication that the particle orientation changes
from parallel to tilted as a
function of the grafting density.

The total excess density at the wall,
\begin{equation}
\rho_{\mbox{\tiny excess}} = \frac{1}{2}\int_0^{L_z} 
dz \: (\rho(z) - \rho_{\mbox{\tiny bulk}}),
\end{equation}
is shown as a function of the grafting density in Figure \ref{fig:dexc}. 
Two opposite effects contribute: Particles are pushed towards the walls, 
but cannot move beyond $z = \sigma_0/2$. At low grafting densities, 
the excluded volume at the wall is not quite compensated by the density 
excess in the density peaks close to the wall; at higher grafting 
densities, it is overcompensated. The overall excess density is small 
and rises roughly linearly with the grafting density. 

\begin{figure}[t]
\noindent
\fig{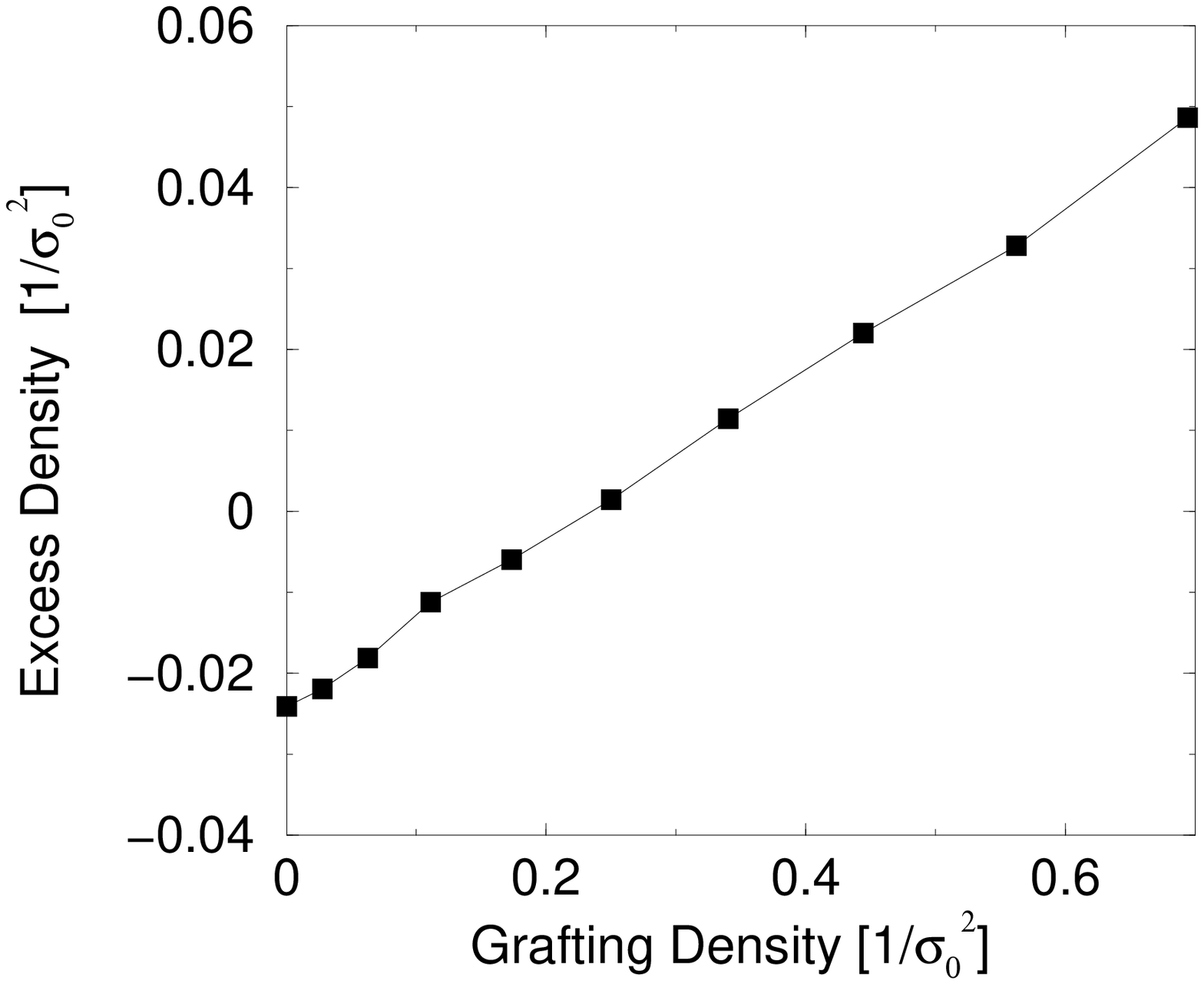}{80}{75}{}
\CCdexc
\end{figure}

\begin{figure}[t]
\noindent
\fig{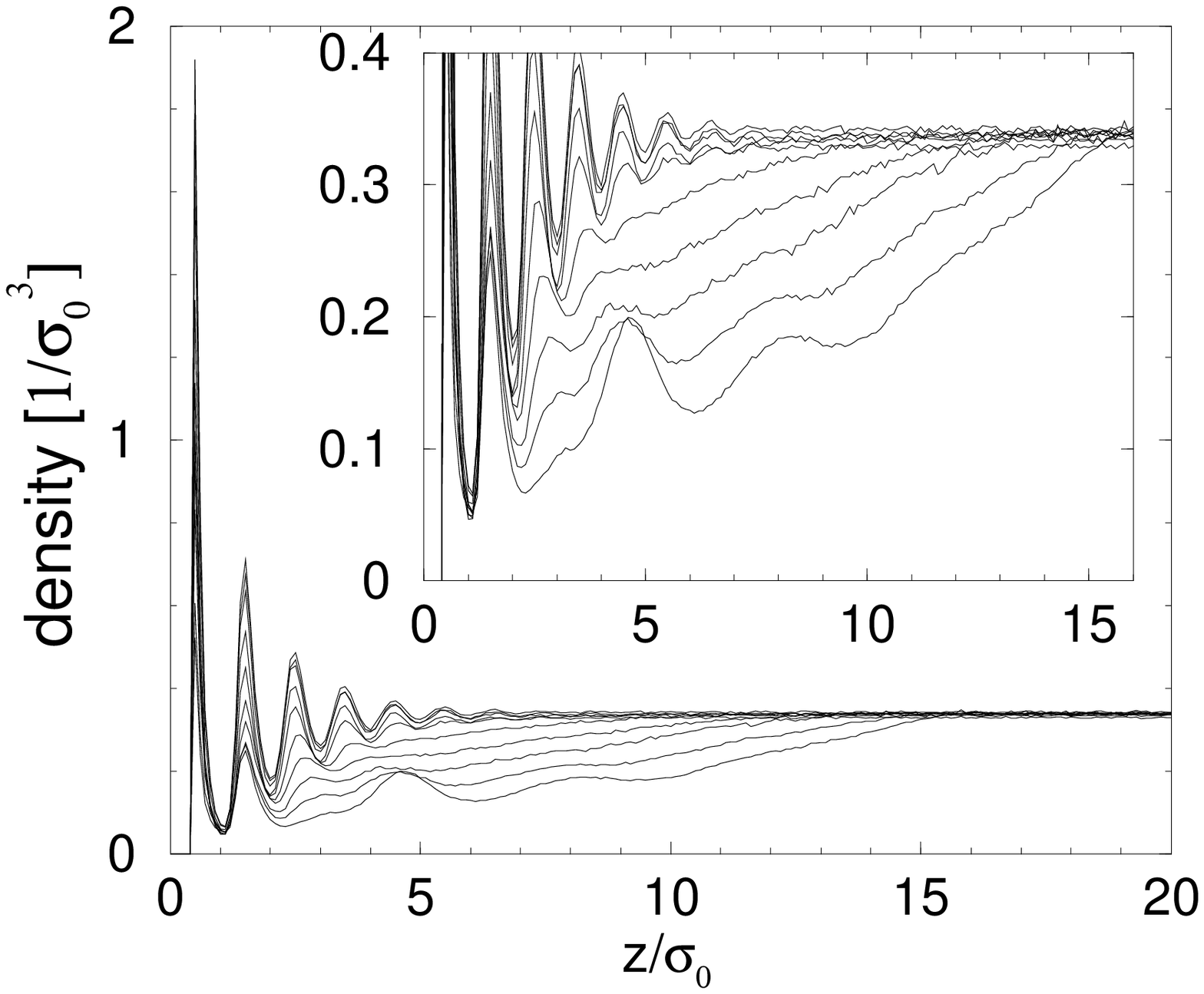}{80}{70}{}
\CCdsolv
\end{figure}

\begin{figure}[t]
\noindent
\fig{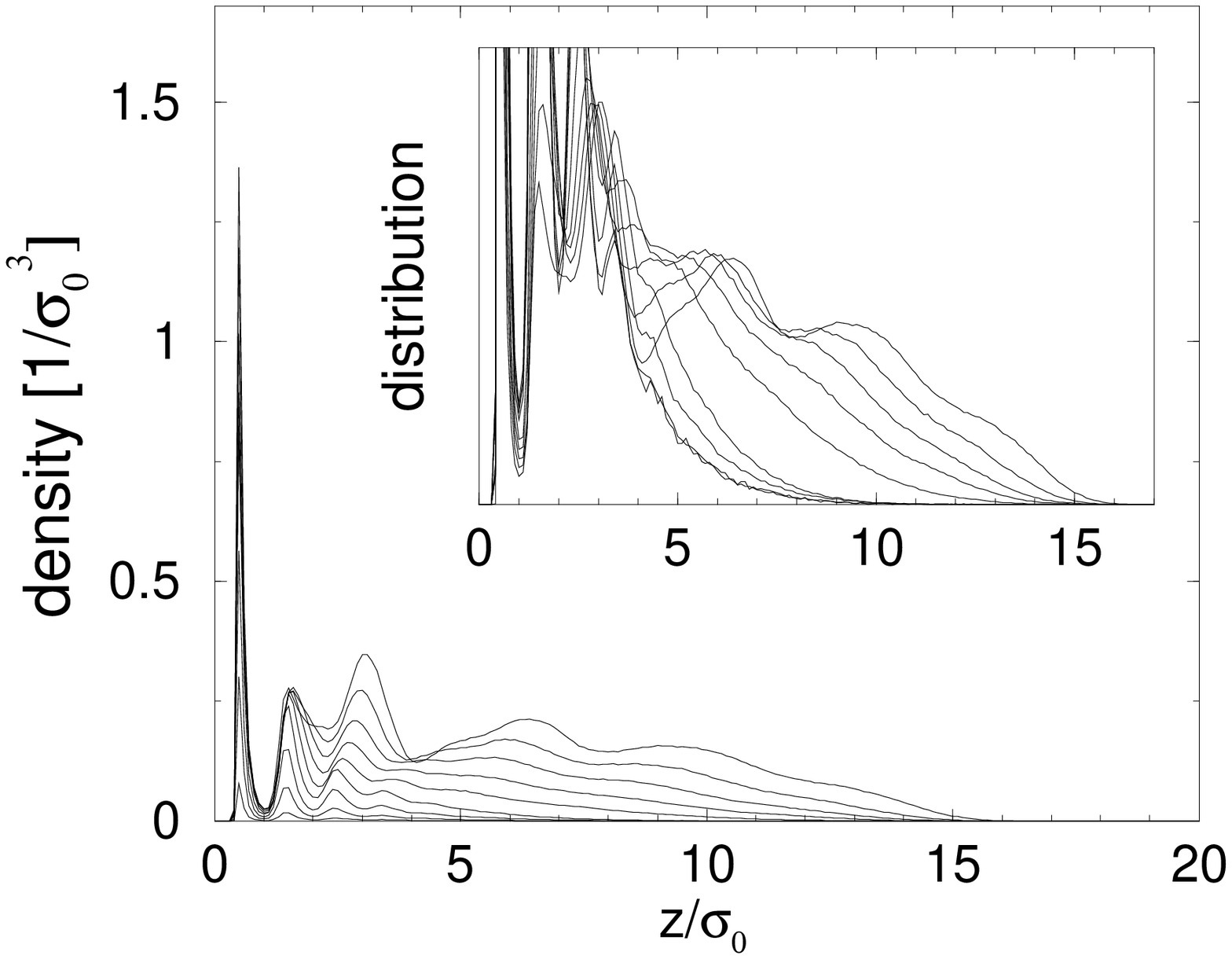}{83}{65}{}
\CCdchain
\end{figure}

It is instructive to analyze the profiles of solvent particles and 
chain monomers separately. The profiles of the solvent particles 
(Figure \ref{fig:dsolv}) show how the solvent particles are successively 
expelled from the chain region as the grafting density increases. The 
profiles of the chain particles are shown in Figure \ref{fig:dchain}. 
Close to the surface, they reproduce the layering of the overall density 
profile.  The form of the profiles far from the surface depends on the 
grafting density: At low grafting densities, they decay smoothly; at higher 
grafting densities, they develop 
\fig{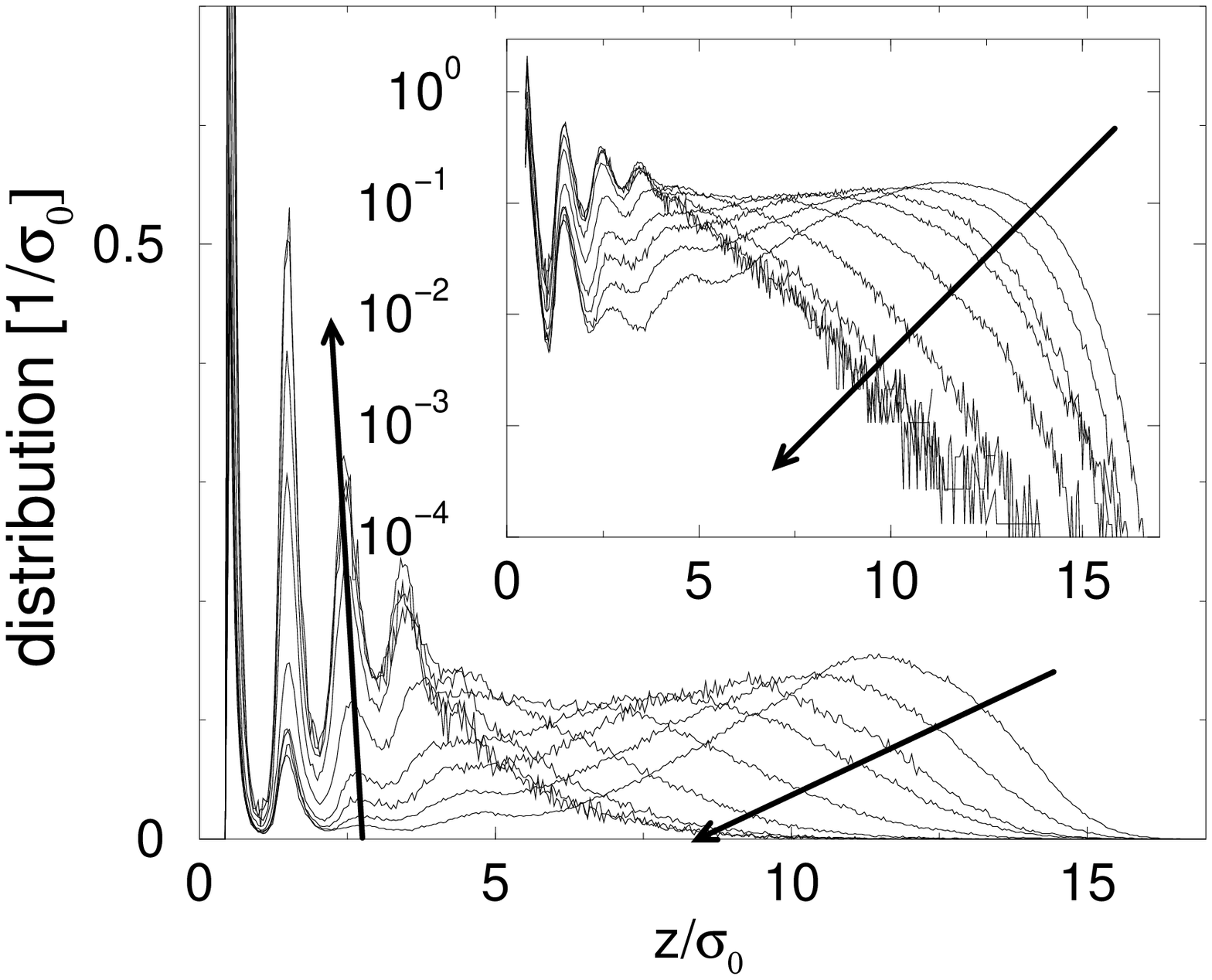}{83}{73}{}
\begin{figure}[t]
\CCdends
\end{figure}

\noindent
structure. Additional insight can be gained 
from looking at the normalized chain density distributions at different 
grafting densities  $\Sigma$ (Figure \ref{fig:dchain}, inset). They are almost 
identical for the lowest $\Sigma$. As the grafting
density increases, more and more chain monomers are
pushed away from the surface, i. e., the chains stand up.

To study this in more detail, we inspect the distributions of chain 
ends in the $z$ direction (Figure \ref{fig:dends}). At low grafting 
densities, the chain end distributions again reproduce the layered 
structure of the nematic fluid close to the wall and decay smoothly,
almost exponentially, far from the wall. At the grafting density 
$\Sigma/\sigma_0^2 \sim 0.15$, a new broad peak emerges at some 
distance from the wall, which grows and moves away from the wall as 
the grafting density increases further. However, a fraction of
chain ends still remains in the peaks close to the surface.
Thus one observes two ``types'' of chains: Some that lie flat on the 
surface, and some that gradually stand up. Increasing the grafting 
density affects the conformations of the chains in two ways: 
The fraction of chains that lie flat decreases, and the chains
that stand up extend deeper into the bulk.

Next we study the tilt angle $\theta$ of chains and particles with 
respect to the surface normal. Figure \ref{fig:tchains} shows the 
average angle $\langle \theta \rangle$ of the head-to-tail vector of 
chains and compares it with the corresponding value in a ``dry'' 
system of grafted chains without solvent particles. The dry layer 
is a reference system of randomly oriented chains, which are only 
subject to their mutual repulsive interactions and to the repulsive 
interaction with the wall. In the presence of solvent, the tilt angle 
of the chains is mostly higher than in the absence of solvent. At low 
\fig{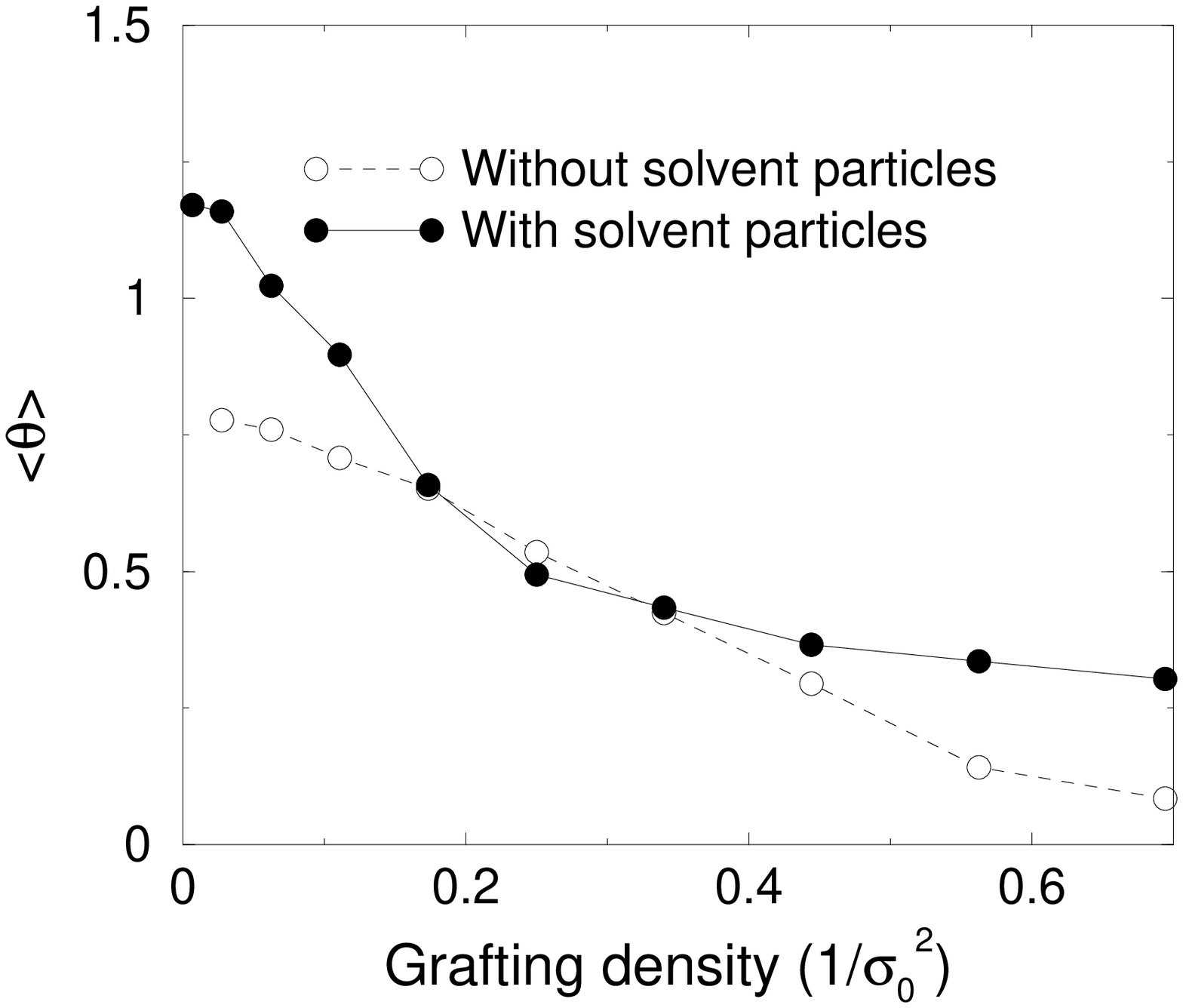}{75}{70}{}
\noindent
\begin{figure}
\CCtchains
\end{figure}

\noindent
grafting densities, the nematic solvent determines the director profile 
close to the surface, and the chains lie down, i.e., $\theta$ is large. 
As the grafting density increases, the tilt angle drops rapidly and 
reaches that of the reference system at $\Sigma \sim 0.2 /\sigma_0^2$. 
Beyond $\Sigma \sim 0.3/\sigma_0^2$, the tilt angle almost ceases to 
decrease. The chains retain tilt even at grafting densities where the 
chains in the dry system are almost perpendicular. 

Compared to the dry system, the nematic solvent introduces a second 
important feature: It breaks the azimuthal symmetry. In the absence
of solvent, the chains are disordered. In the presence of solvent,
they tilt collectively in one direction. The direction of the tilt 
is related to the local direction of the director.

\begin{figure}[t]
\noindent
\fig{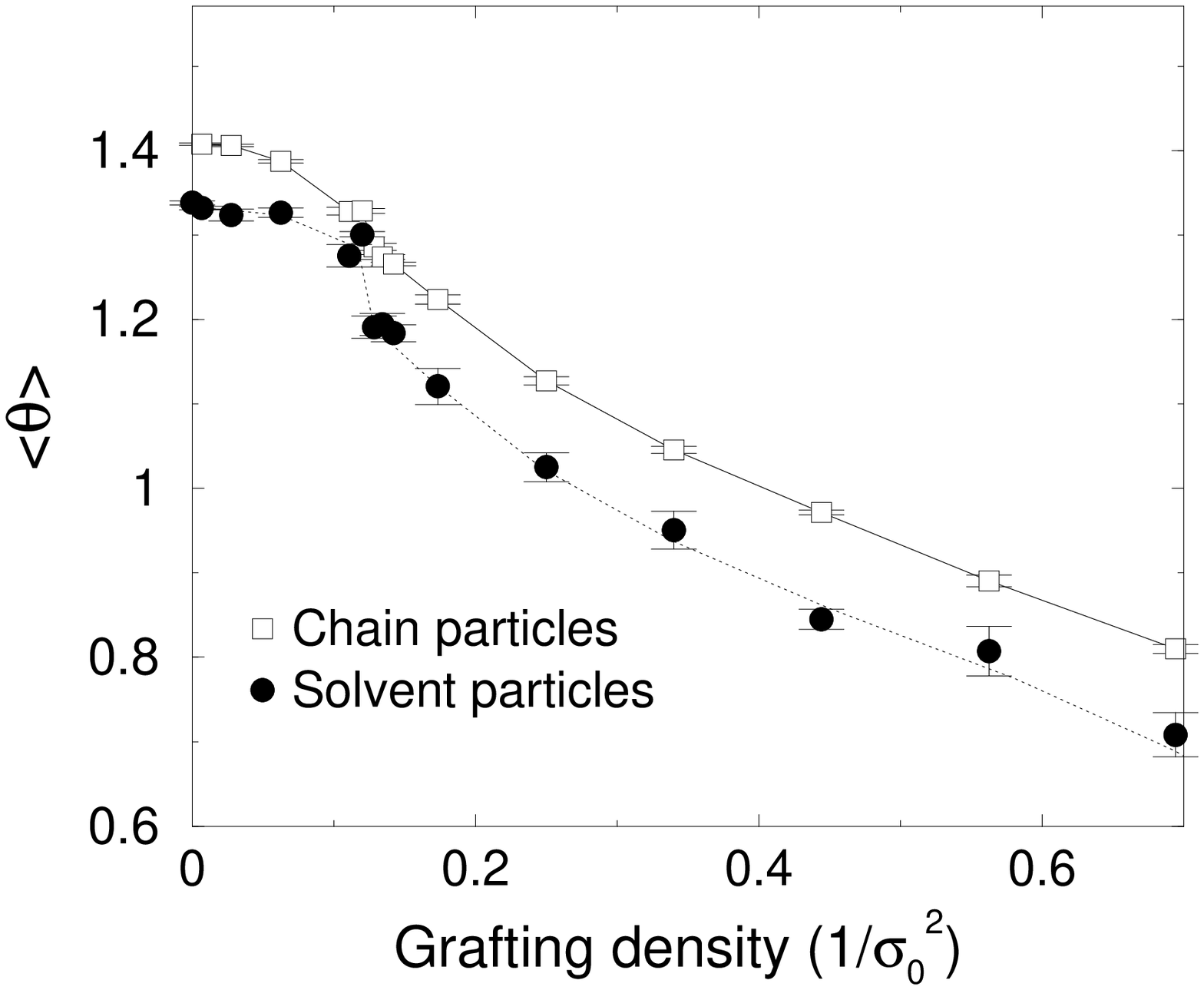}{75}{70}{}
\CCtparticles
\end{figure}

After having studied the tilt of whole chains, we discuss the tilt of 
individual particles. The average angle $\langle \theta \rangle$ for chain 
monomers and solvent particles is shown as a function of the grafting density 
in Figure \ref{fig:tparticles}. The average tilt of chain monomers is
consistently higher than that solvent particles, i. e., the chain region
is more tilted than the bulk of the system. Looking at the average
tilt of the solvent particles, one notices a a small jump at the grafting 
density $\Sigma^* \approx 0.13/\sigma_0^2$. This suggests the presence 
of a first order phase transition. At grafting densities lower than
$\Sigma^*$, the particles lie almost parallel to the wall. Beyond 
$\Sigma^*$, they exhibit tilt.

We can study these phenomena in more detail by inspecting the distribution 
$P(\cos(\theta))$ of tilt angles for solvent particles and chain particles,
shown in in Figure \ref{fig:thist}. It is normalized such that the average 
number of particles in an angle window $[\theta,\theta+d\theta]$ is given by
$N_{\theta, d\theta} = N_{total} P(\cos (\theta)) \sin(\theta) d \theta $.
At low grafting densities, the distribution has always its maximum at 
$\theta = \pi/2$, i.e., the particles lie flat at the wall. If one 
increases the grafting density, the distribution first broadens without 
shifting. At the grafting density $\Sigma \sim 0.13/\sigma_0^2$, the maximum 
of the distribution detaches from $\pi/2$ rather abruptly and starts moving 
to lower tilt angles. From then on, the maximum tilt angle decreases 
continuously with increasing grafting density. The chain monomer 
distribution has a small secondary maximum at the angle $\pi/2$ at 
all grafting densities. This is the contribution of the fraction
of chains which are flat on the substrate.

The maximum angle $\theta{\mbox{\tiny max}}$ is plotted as a function of 
the grafting density in Figure \ref{fig:thistmax}. The first order phase
\fig{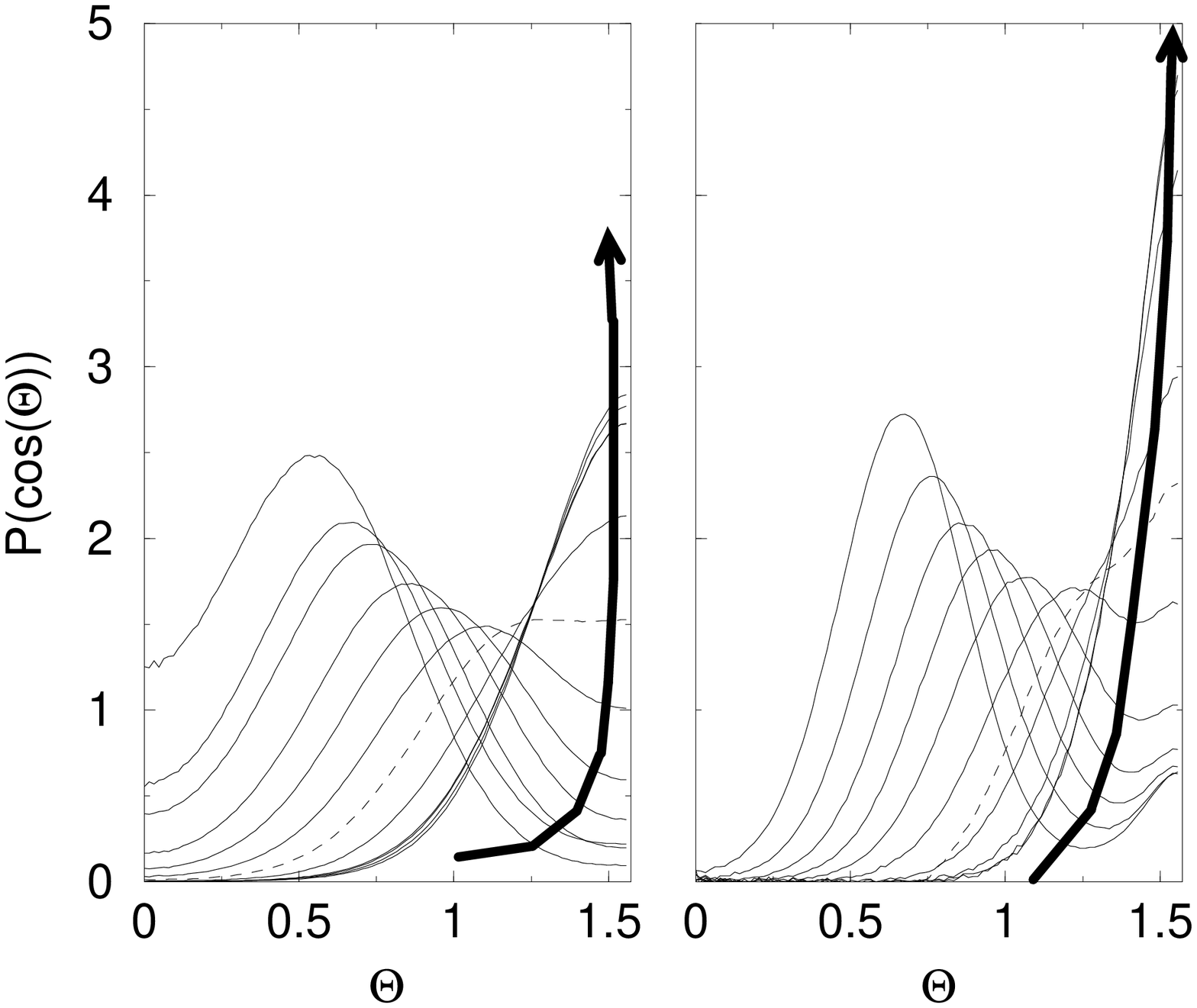}{85}{75}{}
\begin{figure}[t]
\CCthist
\end{figure}

\begin{figure}[t]
\noindent
\fig{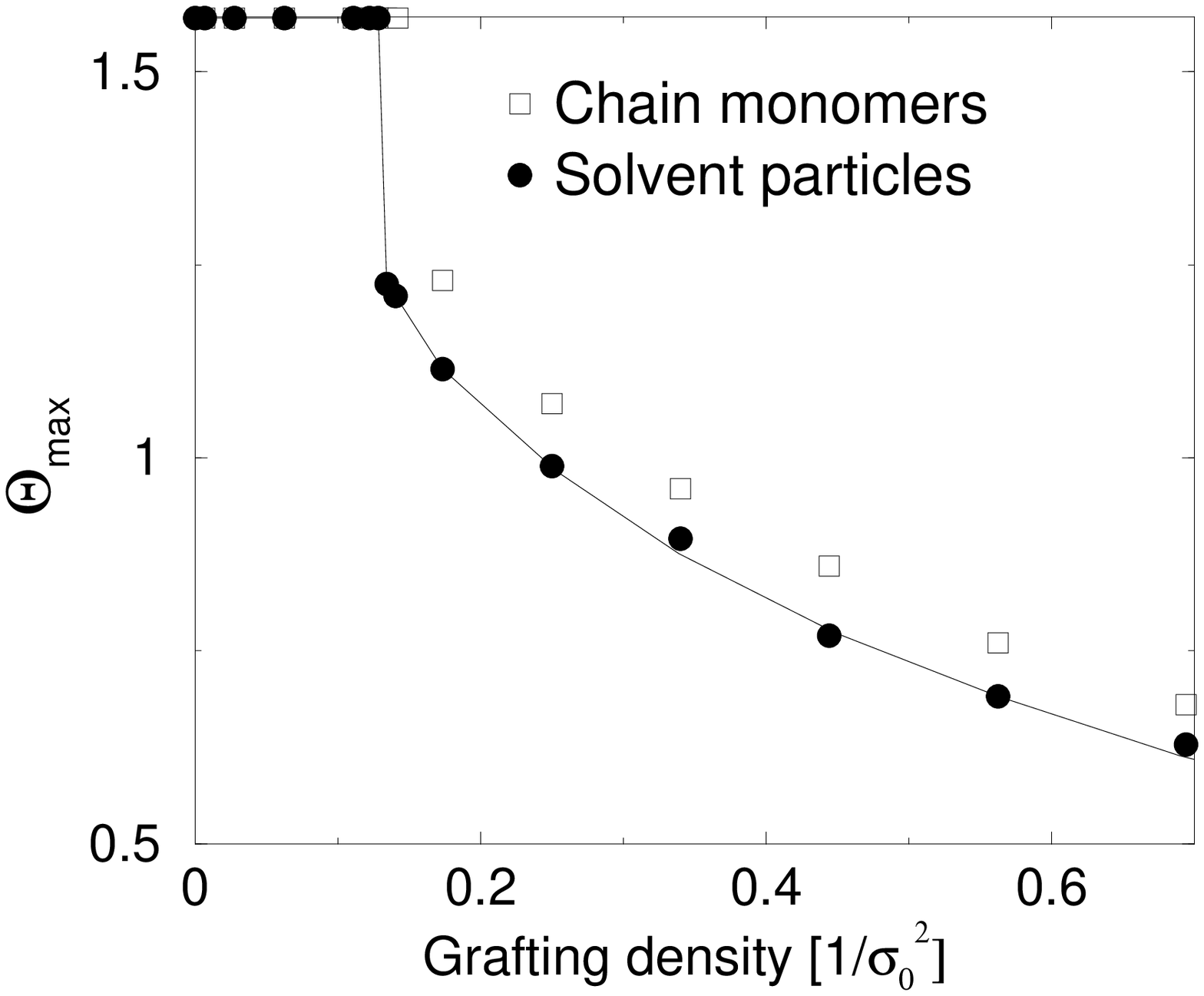}{70}{65}{}
\CCthistmax
\end{figure}

\noindent
transition in the nematic film becomes now very evident. The state of
the film is best characterized by the angle distribution of the solvent
particles, because the majority of them is outside of the chain layer.
In contrast, the chain monomer distribution characterizes the structure 
close to the surface. The angle $\theta{\mbox{\tiny max}}$ for solvent 
particles jumps to a value different from $\pi/2$ at 
$\Sigma^* = 0.13/\sigma_0^2$. In the case of the chain particles, 
the $\theta_{\mbox{\tiny max}}$ jumps at a slightly higher grafting 
density ($\Sigma = 0.17/\sigma_0^2$): The chains which lie 
flat strengthen the peak at $\theta = \pi/2$. 

Finally, we turn to the discussion of the director profiles. The nematic 
order parameter and the nematic director in a slice of the simulation box 
are defined as follows: From the orientations $\uu_i$ of all $n$ particles 
$i$ in the slice, one calculates the order tensor, 
\begin{equation}
\label{eq:op}
\QQ = \frac{1}{n} \sum_{i=1}^{n}
(\frac{3}{2} \uu_i \otimes \uu_i - \frac{1}{2} {\II} ) ,
\end{equation}
where $\II$ denotes the unity matrix and $\otimes$ the dyadic product. The 
largest eigenvalue of this matrix is the order parameter $S$ in the slice, 
and the corresponding eigenvector is the director $\nn$. The order parameter 
in a bulk system with our parameters is roughly $S = 0.75$. By construction, 
$S$ increases systematically if evaluated in subsystems with few particles. 
Therefore it tends to appear larger in a thin slice than in the bulk.  

In order to calculate order parameter and director profiles, we cut the 
system into regular slices of thickness $\Delta z=1 \sigma_0$, starting 
from the left and the right wall up to almost the middle of the film. 
A thicker slice in the middle collected the remaining particles and absorbed 
the fluctuations of the box length $L_z$. This slice was then disregarded.

\begin{figure}[t]
\noindent
\fig{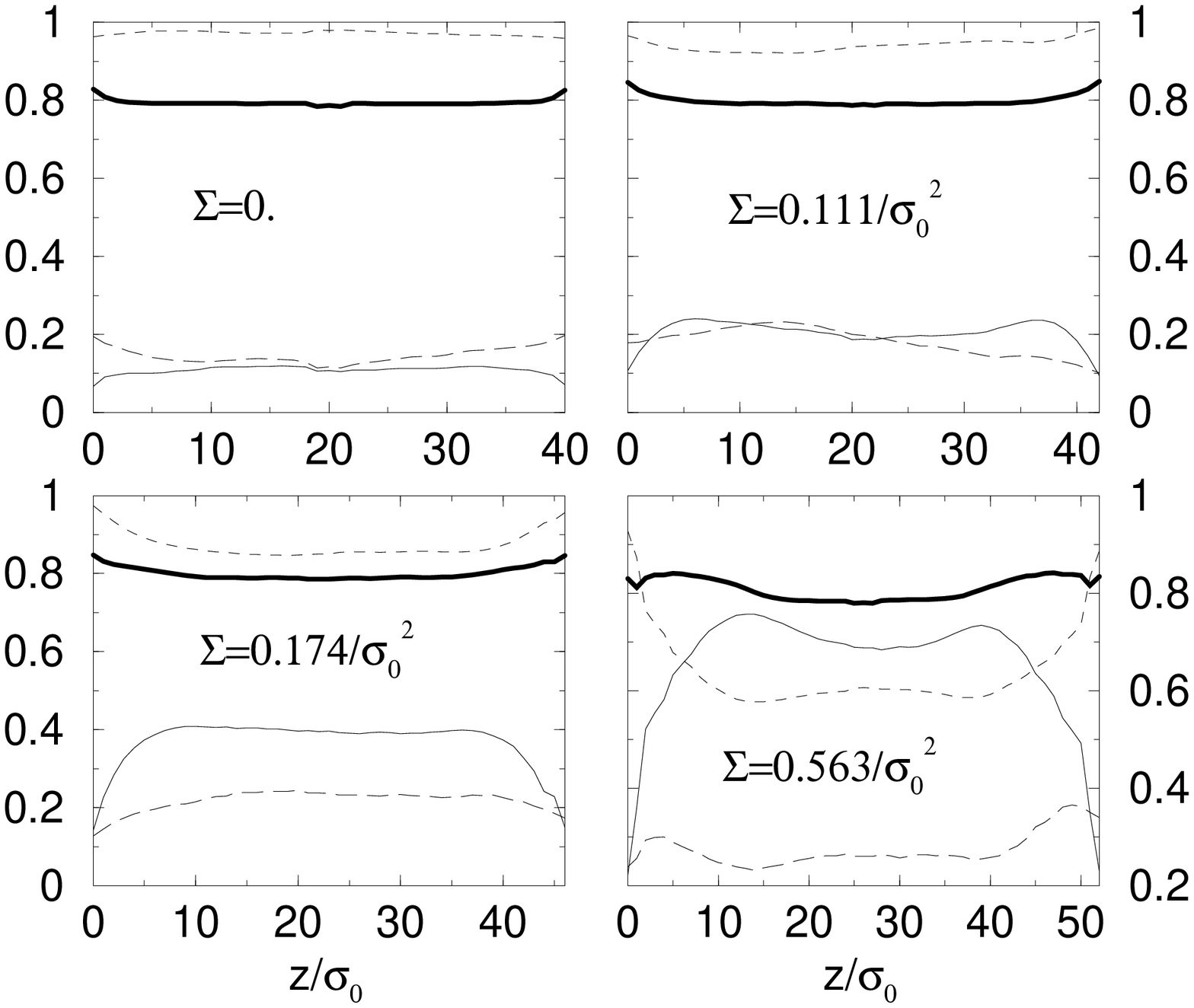}{85}{75}{}
\CCprofop
\end{figure}

Results for four grafting densities are shown in Figure \ref{fig:profop}. 
The order parameter $S$ varies very little throughout the film. In contrast, 
the director $\nn$ changes quite dramatically at higher grafting densities: 
Close to the walls, it is always oriented parallel to the wall, $n_z \ll 1$. 
At low grafting densities, it stays parallel everywhere. At high grafting 
density, it turns around very rapidly within the first few $\sigma_0$ of 
the surface, and then continues to turn more slowly. Substantial 
reorientation takes place in a $z$-region of about the thickness of the 
chain layer (cf. Figure \ref{fig:dchain}). As a result, the director
is tilted in the middle of the film.

The $y$-component of the director, $\langle |n_y| \rangle$, is overall 
smaller than the $x$-component. This is because the systems were initially 
prepared such that the director pointed in the $x$ direction. The diffusion 
time scale for reorientation without any driving force is much longer than 
the lengths of our runs (several million Monte Carlo steps). In one case, 
however, we happened to observe the onset of a reorientation -- the 
system with grafting density $\Sigma = 0.00694/\sigma_0^2$ ventured into a 
configuration where the director twisted between the $x$ and the $y$ 
direction after 5 million Monte Carlo steps. The twist disappeared 2 million 
Monte Carlo steps later. It affected the profile of $\langle n_z \rangle$ 
only slightly.

Looking at Figure \ref{fig:profop}, one notices furthermore that the
director profile hardly finds enough space to reach a plateau in the middle 
of the film at high grafting densities. This is a serious problem: Our 
systems of length $L_z \approx 50 \sigma_0 (\pm 10 \sigma_0)$ are obviously 
too small to accommodate a real bulk region between the two surfaces. 
The results are subject to strong finite size effects, and any conclusions 
about the anchoring behavior in a semi-infinite system must be viewed with 
caution. Additional finite size effects can be expected due to the limited 
lateral size of the simulation box, $L_{\parallel} = 12 \sigma_0$. 
It would be desirable to study these effects by varying $L_{\parallel}$ 
and $L_z$ systematically in different runs. Unfortunately, an extensive
finite size analysis is computationally too expensive at the moment.
Hence our quantitative results can be questioned. Nevertheless, we believe 
that the observed effects will persist qualitatively in semi-infinite
systems. 

These caveats stated, we collect the results for the order parameter 
$n_z$ in the middle of the film at different grafting densities 
(Figure \ref{fig:nzfilm}). In a macroscopically thick 
film, it is directly related to the anchoring angle via 
$n_z = \cos(\theta_{\mbox{\tiny anchoring}}$).
Figure \ref{fig:nzfilm} confirms and completes the picture already suggested 
by Figures \ref{fig:tparticles} and \ref{fig:thistmax}: At low grafting 
densities, the surface anchors the director parallel to the wall. 
The director component $\langle | n_z | \rangle $ remains constant 
and close to zero over a range of grafting densities, up to 
$\Sigma^* = 0.13 /\sigma_0^2$. Then it jumps to a nonzero value and 
rises continuously thereafter. The transition from planar to tilted 
alignment is clearly discontinuous: A range of values $n_z$ is ``forbidden''. 
Hence a small range of anchoring angles $\theta{\mbox{\tiny anchoring}}$ 
close to $\theta=\pi/2$ ($\theta \in [0.4 \pi:0.5 \pi]$) cannot by accessed 
at any grafting density. As mentioned above, finite size effects have
to be expected. We cannot exclude the possibility that large scale 
fluctuations drive the transition to second order. But since the quantity 
most susceptible to large scale fluctuations in the tilted phase is 
presumably the azimuthal tilt angle, we suspect that the fluctuations 
will most likely reduce the azimuthal ordering and hence
reduce the anchoring angle $\theta$. This implies that the gap in Figs.
\ref{fig:thistmax} widens, and that the transition becomes more
strongly first order.

\begin{figure}[t]
\noindent
\fig{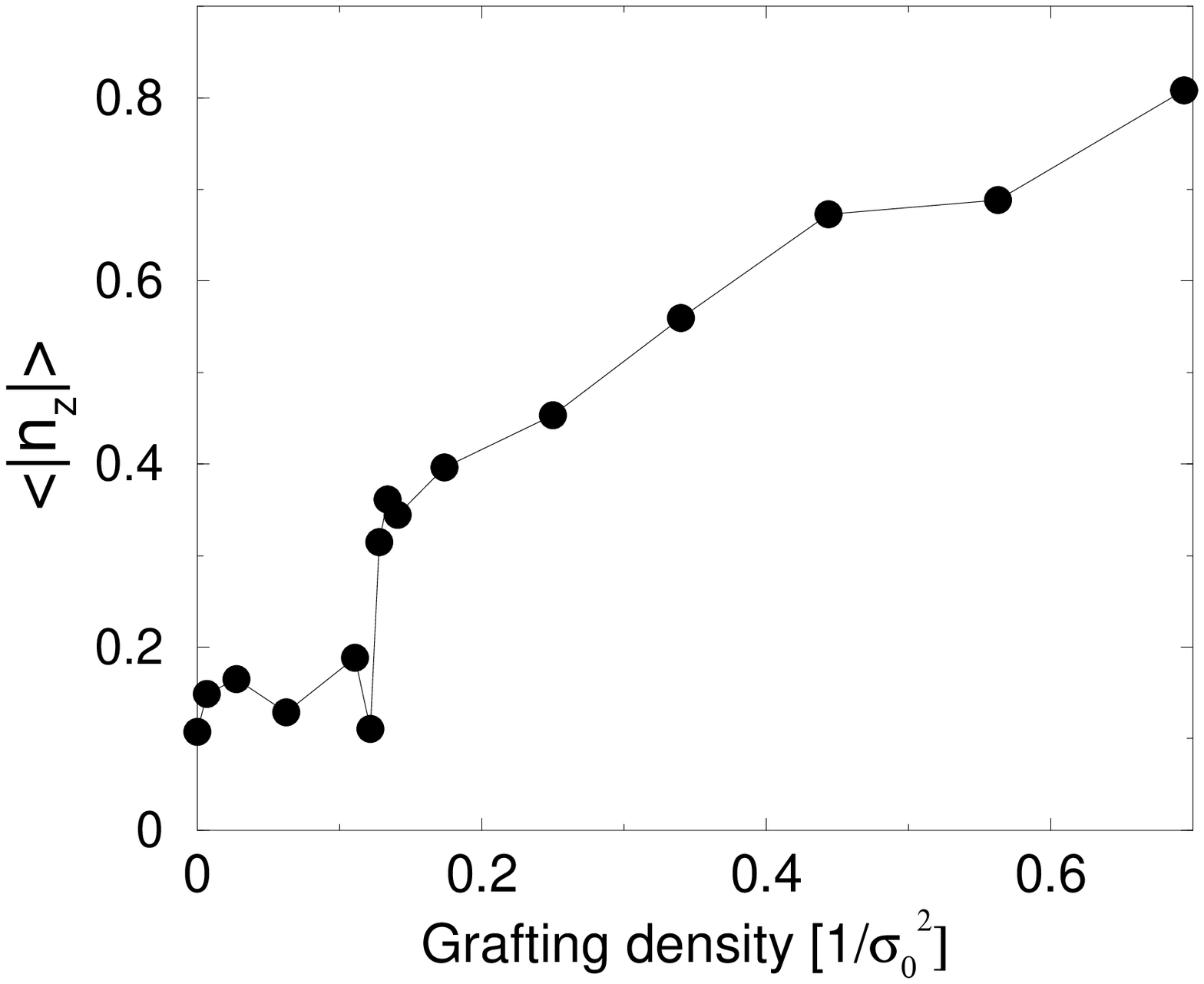}{80}{80}{}
\CCnzfilm
\end{figure}

\section{Summary and Discussion}
\label{sec:summary}

In sum, we have studied the anchoring of a model nematic fluid on layers of
liquid-crystalline chain molecules, grafted on a substrate which favors 
parallel anchoring. At low grafting densities, the substrate dominates 
and the nematic fluid is oriented parallel to the wall. A first order 
transition to a phase with tilted anchoring is encountered at a grafting 
density $\Sigma^*$. On increasing the grafting density further, the director 
gradually stands up. The phase transition goes along with a major 
reorganization within the chain layer: Below $\Sigma^*$, all chains lie 
flat at the wall. Above $\Sigma^*$, the chains separate into two distinct 
classes: Some remain flat at the wall, and some stand up. 
The tilt angle $\theta$ of the chains belonging to the latter group 
decreases with increasing grafting density, and the relative number of 
chains in that group increases.

The transition is very similar to the anchoring transition predicted by 
Halperin and Williams~\cite{avi1}: It mediates between a phase with 
parallel anchoring at low grafting density, $\Sigma < \Sigma^*$, and a 
tilted phase at higher grafting density, $\Sigma > \Sigma^*$. It is driven 
by a competition between the anchoring force from the substrate, which favors 
parallel anchoring, and the repulsive forces between the grafted chains,
which force them to stand up at high grafting densities. However, the 
details of the underlying mechanisms are quite different. The theory of 
Halperin and Williams is devised for long chains with many hairpins, which 
can be treated as (anisotropic) entropic springs. Our chains are too short 
to support hairpins, and the configurational entropy is much less important. 
The other two constituents of the theory of Halperin and Williams, the 
excluded volume of the chains and the elastic energy of the nematic solvent, 
are obviously sufficient to bring about the phase transition. 

The simulations and theories~\cite{avi1,avi2,avi3,harald1} consider anchoring 
on layers of main-chain liquid crystalline chains. Experimentally, layers of
grafted side-chain liquid crystalline polymers can be synthesized much
more easily~\cite{peng2,peng3}. In these molecules, the liquid crystalline 
blocks orient themselves preferably perpendicular to the backbone of the 
chains. In order to generate a conflict between chain stretching and the 
anchoring on the substrate, one needs to prepare the substrate such that it 
aligns the liquid crystals in a homeotropic way. So far, this has not been 
done systematically, but preliminary studies by Peng et al show that it 
should be feasible~\cite{peng3}. The question is: can we expect that an effect
which has been predicted for main-chain liquid crystalline polymers
will be observed in an experiment with side-chain liquid crystalline
polymers? 

Our study might help to answer this question. Comparing our simulation 
results with the theory of Halperin and Williams, we can conclude that the 
details of the chain conformations and the chain entropy do not really matter 
for the anchoring transition. It is robust enough to survive even in a regime 
for which the theory was never designed. Therefore the chances that it can 
be observed in real systems should be quite high.

\begin{acknowledgments}

We have benefitted from useful conversations with A. Halperin,
D. Johannsmann, J. R\"uhe, B. Peng, K. Binder, and in
particular with M. P. Allen.  We thank K. Binder and M. P. Allen
for allowing us to perform simulations on the computers
of their groups in Bristol and Mainz. This work was funded
by the German Science Foundation (DFG).

\end{acknowledgments}

\end{document}